\documentclass[amsmath,amssymb,twocolumn,superscriptaddress]{revtex4}
\usepackage{aasjournals}
\usepackage{graphicx}%
\usepackage{dcolumn}%
\def\be{\begin{equation}}
\def\ee{\end{equation}}
\def\beq{\begin{eqnarray}}
\def\eeq{\end{eqnarray}}

\usepackage{bm}
\usepackage{graphicx}
\usepackage{dcolumn}
\usepackage{amsmath}
\usepackage{xcolor}
\usepackage{graphicx, psfrag}
\usepackage{longtable}
\usepackage{amssymb}
\usepackage[colorlinks=true, citecolor=blue, urlcolor = blue, linkcolor= red, bookmarks=true]{hyperref}
\usepackage{float}
\usepackage{mathtools}
\usepackage{amsmath}
\usepackage{amsfonts}
\usepackage{dcolumn}
\usepackage{hyperref}
\usepackage{subfigure}
\usepackage{pgfplots}
\usepackage{epstopdf}
\usepackage{booktabs}
\usepackage{orcidlink}
\usepackage{xcolor}
\usepackage{booktabs}
\usepackage{listings}

\begin{document}
\title{Probing Cosmic Curvature with Fast Radio Bursts and DESI DR2}

\author{Jéferson A. S. Fortunato \footnote{Corresponding author} \orcidlink{0000-0001-7983-1891} }
\email[]{jeferson.fortunato@edu.ufes.br}
 \affiliation{PPGCosmo, CCE, Universidade Federal do Esp\'{\i}rito Santo (UFES), Av. Fernando Ferrari, 540, CEP 29.075-910, Vit\'oria, ES, Brazil}
 \affiliation{Instituto Argentino de Radioastronom\'{\i}a, CCT La Plata, CONICET; CICPBA; UNLP, C.C.5, (1894) Villa Elisa, Argentina}
 \affiliation{High Energy Physics, Cosmology \& Astrophysics Theory (HEPCAT) Group, Department of Mathematics and Applied Mathematics, University of Cape Town,
Cape Town 7700, South Africa}

\author{Wiliam S. Hip\'olito-Ricaldi \orcidlink{0000-0002-1748-553X}
}
\email[]{wiliam.ricaldi@ufes.br}
\affiliation{Departamento de Ci\^encias Naturais, CEUNES, Universidade Federal do Esp\'{\i}rito Santo (UFES), Rodovia BR 101 Norte, km. 60,\\
CEP 29.932-540, S\~ao Mateus, ES, Brazil}
\affiliation{N\'ucleo Cosmo-UFES, CCE, Universidade Federal do Esp\'{\i}rito Santo (UFES), Av. Fernando Ferrari, 540, CEP 29.075-910, Vit\'oria, ES, Brazil}

\author{Gustavo E. Romero \orcidlink{0000-0002-5260-1807}
}
\email[]{romero@iar.unlp.edu.ar}
\affiliation{Instituto Argentino de Radioastronom\'{\i}a, CCT La Plata, CONICET; CICPBA; UNLP, C.C.5, (1894) Villa Elisa, Argentina}
\affiliation{Facultad de Ciencias Astron\'{\o}micas y Geof\'{\i}sicas, Universidad Nacional de La Plata,
B1900FWA La Plata, Argentina}

\begin{abstract}\noindent
The spatial curvature of the Universe remains a central question in modern cosmology. In this work, we explore the potential of localized Fast Radio Bursts (FRBs) as a novel tool to constrain the cosmic curvature parameter $\Omega_k$ in a cosmological model–independent way. Using a sample of 120 FRBs with known redshifts and dispersion measures, we reconstruct the Hubble parameter $H(z)$ via artificial neural networks, and use it to obtain angular–diameter distances $D_A(z)$ through two complementary approaches. First, we derive the comoving distance $D_C(z)$ and $D_A(z)$ directly from FRBs without assuming a fiducial cosmology. Then, we combine the FRB-based $H(z)$ with Baryon Acoustic Oscillation (BAO) DESI DR2 measurements to infer $D_A(z)$.\@
By comparing the FRB-derived and BAO+FRB-derived $D_A(z)$, we constrain spatial curvature. Our covariance-based likelihood (accounting for correlated uncertainties) yields $\Omega_k = -0.31\pm0.57$, while a diagonal (Gaussian) treatment gives $\Omega_k = -0.13\pm0.46$. Both estimations are consistent with spatial flatness at the $1\sigma$ level, albeit with a mild preference for negative curvature. Explicitly accounting for the full covariance broadens the intervals and avoids underestimation of uncertainties. These results highlight the growing relevance of FRBs in precision cosmology and their synergy with BAO as a powerful, cosmological model–independent probe of the large–scale geometry of the Universe.
\end{abstract}

\maketitle

\section{Introduction}
Determining the spatial geometry of the Universe through the curvature parameter $\Omega_k$ is a fundamental goal in cosmology. It directly impacts the measurement of distances, as well as our understanding of spacetime, the global topology of the Universe, its ultimate fate, and the physics of the early times, for instance. 

The current constraints on $\Omega_k$ are largely consistent with flatness, although uncertainties remain. For instance, combining \textit{Planck} Cosmic Microwave Background (CMB) lensing data with low-redshift Baryon Acoustic Oscillations (BAO) measurements yields  $\Omega_k = 0.0007 \pm 0.0019$ \cite{aghanim2020planck}. However, other analyses challenge this result. Notably, \cite{di2020planck} reported a preference for a mildly closed Universe, with $\Omega_k = -0.044^{+0.018}_{-0.015}$, based solely on CMB temperature and polarization data from \textit{Planck} satellite. Another study found a preference for a closed Universe at the $2\sigma$ confidence level, measuring $\Omega_k = -0.089^{+0.049}_{-0.046}$ when applying Effective Field Theories of Large Scale Structure (EFTofLSS) to BAO data \cite{glanville2022full}. On the other hand, \cite{wu2024measuring} reported $\Omega_k=0.108\pm0.056$, which is in $2.6\sigma$ tension with the CMB lensing data result. To obtain this estimate, they used a combination of datasets, including BAO, Cosmic Chronometers (CC), and Strong Gravitational Lensing (SGL). Other recent studies reached conclusions consistent with spatial flatness by combining alternative datasets, reporting mean values of $\Omega_k$ that deviate slightly from zero but remain within $1\sigma$ of flatness. For instance, in \cite{vagnozzi2021galaxy} used full-shape galaxy power spectra from BOSS DR12 CMASS and \textit{Planck} data, finding $\Omega_k = 0.0023 \pm 0.0028$, while \cite{vagnozzi2021eppur} combined \textit{Planck} with CC data to obtain $\Omega_k = -0.0054 \pm 0.0055$.

However, it is important to recognize that these constraints are inherently cosmological model-dependent, relying on assumptions within the non-flat $\Lambda$CDM framework, which suffers from the well-known intrinsic degeneracy between cosmic curvature and the Hubble constant. Moreover, as with the Hubble tension, the apparent disagreement in curvature estimates raises the question of whether systematic biases or new physics are at play. In this context, exploring methods that provide a complementary and cosmological model-independent perspective is particularly valuable. Developing curvature measurements that are independent of the cosmological model, especially those that do not rely solely on CMB-based assumptions, offers a promising pathway to gaining a better understanding of these issues. Recent efforts have pursued this direction using non-parametric reconstructions with combinations of SNIa, BAO, and CC data, providing curvature constraints at the $\mathcal{O}(10^{-1})$--$\mathcal{O}(10^{-2})$ level without relying any specific expansion history priors \cite{dhawan2021non, jiang2024nonparametric}.

Recent studies \cite{gong2024multiple,liu2025newest}, which provide a methodological basis for the present work, have proposed model-independent approaches by combining BAO and Hubble parameter data. Although their methodologies are broadly similar, the former utilizes only BOSS and eBOSS BAO data and omits the covariance matrix, whereas the latter incorporates full covariance information and includes additional measurements from DESI DR1. Both studies find results consistent with flat spatial curvature within the reported uncertainties. 

Extending these methodologies, we adopt a comparable framework to systematically assess the impact of including versus omitting the covariance matrix on cosmic curvature estimates. Additionally,  we incorporate FRBs as novel and independent tracers to reconstruct the Hubble parameter $H(z)$, offering a complementary perspective to traditional probes such as supernovae or cosmic chronometers. This inclusion introduces a new observational channel based on dispersion measures, thereby enhancing the robustness and versatility of model-independent curvature constraints. We further combine this with the latest BAO data from BOSS, eBOSS, and DESI DR2 \cite{karim2025desi}.

FRBs have emerged as a promising new probe in cosmology. These events are characterized by millisecond-duration and highly luminous pulses in the radio frequency spectrum. First discovered in 2007 by Lorimer et al.~\cite{lorimer2007bright}, FRBs exhibit large dispersion measures (DMs), that significantly exceed the expected contribution from the Milky Way. This excess provides strong evidence for their extragalactic, and potentially cosmological, origin \cite{petroff2019fast}. Since that initial detection, hundreds of FRBs have been observed, and a subset has been precisely localized to host galaxies with measured redshifts \cite{zhou2022fast}.

Beyond their intriguing astrophysical nature, FRBs have proven to be valuable probes for cosmology. In particular, localized FRBs, those with measured redshifts and host galaxy associated, provide unique observational access to the otherwise elusive ionized content of the intergalactic medium (IGM). Over the past few years, multiple studies have employed these events to estimate key cosmological quantities, such as the Hubble constant $H_0$ \cite{macquart2020census, wu20228, zhao2022first, kalita2025fast, hagstotz2022new,  fortunato2025fast}, cosmographic parameters \cite{fortunato2023cosmography}, the cosmic proper distance \cite{yu2017measuring}, the baryon fraction in the IGM \cite{li2020cosmology, yang2022finding, lemos2022model}, constrain dark energy equation of state \cite{Zhou2014,Gao2014}, trace the magnetic fields in the IGM \cite{Akahori2016}, probe the equivalence principle \cite{Wei2015,Tingay2016,Nusser2016}, the rest mass of photons \cite{wu2016,shao2017,Lin2023}, to cite some. 

In this work, we implement a method to determine the cosmic curvature by leveraging the synergy between FRBs and BAO measurements, with FRBs serving as the core observable. By comparing two estimates of the angular diameter distance \( D_A(z) \), independently reconstructed from FRBs alone and from FRBs+BAO data sets, we infer $\Omega_k$ without relying on a specific cosmological model. This is made possible by using the FRB dispersion measure to reconstruct $H(z)$ through an artificial neural network algorithm, as demonstrated in \cite{fortunato2025fast} and to trace the comoving distance $D_C$ via the integrated electron density along the line of sight.

We assume the validity of the Cosmological Principle, adopting the premise that the Universe is homogeneous and isotropic on large scales. Consequently, the background spacetime is described by the Friedmann–Lemaître–Robertson–Walker (FLRW) metric, with the curvature parameter \(\Omega_k\) determining the geometry of the Universe: flat (\(\Omega_k = 0\)), open (\(\Omega_k > 0\)), or closed (\(\Omega_k < 0\)).

Throughout this work, the expression \emph{model-independent} denotes that no early-universe calibration or specific cosmological model is assumed: BAO enter only through
dimensionless ratios whose combination cancels $r_d$, and $H(z)$ is reconstructed
directly from FRB data via the Macquart relation. This approach relies on Eqs.~(\ref{DM_igm}) and (\ref{ioniza}) (defined in Sec. \ref{basics}), which connect the FRB dispersion measures to $H(z)$ via minimal physical assumptions (FLRW geometry and standard ionization fractions) without assuming a particular cosmological model. Curvature effects enter explicitly only in the FRB-only through the $S_k$ function mapping from $D_C$ to $D_M$.

This paper is organized as follows:  Section \ref{basics} describes some basic properties of FRBs. 
Section \ref{distances} presents an overview of distance calculations in cosmology within the FLRW framework. In Section \ref{sec4} we discuss the methodology used to reconstruct $H(z)$ and the comoving distance from FRB data, which are subsequently employed to infer the cosmic curvature density parameter. The data utilized in this work are detailed in Section \ref{data}.  
Our results and discussion are presented in Section \ref{results}. Finally, Section \ref{conclu} contains our conclusions.
\section{Basic properties of Fast Radio Bursts}\label{basics}
\label{sec2}
The observed dispersion measure, $\mathrm{DM}_{\mathrm{obs}}$, is the sum of local and extragalactic contributions, expressed as:

\begin{equation}\label{dmobs}
    \mathrm{DM}_{\mathrm{obs}} = \mathrm{DM}_{\mathrm{local}} + \mathrm{DM}_{\mathrm{EG}}(z),
\end{equation}

\noindent where the local component includes contributions from the Milky Way:

\begin{equation}
    \mathrm{DM}_{\mathrm{local}} = \mathrm{DM}_{\mathrm{ISM}} + \mathrm{DM}_{\mathrm{halo}},
\end{equation}

\noindent and the extragalactic component includes contributions from the intergalactic medium (IGM) and the host galaxy of FRBs:

\begin{equation}
    \mathrm{DM}_{\mathrm{EG}} = \mathrm{DM}_{\mathrm{IGM}} + \frac{\mathrm{DM}_{\mathrm{host}}}{(1+z)}.
\end{equation}

Each term in this decomposition corresponds to distinct astrophysical environments. $\mathrm{DM}_{\mathrm{ISM}}$ accounts for free electrons in the interstellar medium of the Milky Way and is commonly estimated using galactic electron density models like NE2001 \cite{cordes2002ne2001} and YMW16 \cite{yao2017new}. We adopt NE2001, as recent studies suggest that YMW16 may overestimate $\mathrm{DM}_{\mathrm{ISM}}$ at low Galactic latitudes \cite{koch2021}. The Galactic halo contribution, $\mathrm{DM}_{\mathrm{halo}}$, is less constrained but estimated in the range $50 - 100~\mathrm{pc}\,\mathrm{cm}^{-3}$ \cite{prochaska2019probing}, based on observational constraints, such as the dispersion measure to the Large Magellanic Cloud (LMC), high-velocity cloud dynamics, and hydrostatic equilibrium models of the galactic halo gas. We then adopt a conservative value of $\mathrm{DM}_{\mathrm{halo}} = 50~\mathrm{pc}\,\mathrm{cm}^{-3}$.

The  intergalactic medium (IGM) dispersion measure, $\mathrm{DM}_{\mathrm{IGM}}$, dominates the observed $\mathrm{DM}_{\mathrm{obs}}$ and is crucial for cosmology. However, $\mathrm{DM}_{\mathrm{IGM}}$ exhibits significant scatter due to electron distribution inhomogeneities along the line of sight. Its mean value is given by \cite{deng2014cosmological}:
\begin{equation}\label{DM_igm}
\langle \mathrm{DM}_{\mathrm{IGM}} \rangle = \left(\frac{3c}{8\pi G m_p}\right) \Omega_b H_0 \int_0^z \frac{(1+z') f_{\mathrm{IGM}}(z') f_e(z')}{E(z')} dz',
\end{equation}

\noindent where $c$ is the speed of light, $G$  the gravitational constant, and $m_p$  the proton mass. $\Omega_b$ is the cosmic baryon density, and $E(z) = H(z)/H_0$ the normalized Hubble parameter. The term $f_{\mathrm{IGM}}(z)$ denotes the baryon fraction in the intergalactic medium, which we take as  $f_{\mathrm{IGM}} = 0.82\pm0.04$ \cite{shull2012baryon, zhou2014fast}. The factor $f_e(z)$ describes the  baryons ionization state and is defined as:
\begin{equation} \label{ioniza}
    f_e(z) = Y_H X_{e,H}(z) + \frac{1}{2} Y_{\mathrm{He}} X_{e,\mathrm{He}}(z),
\end{equation}

\noindent where $Y_H = 0.75$ and $Y_{\mathrm{He}} = 0.25$ are the mass fractions of hydrogen and helium, respectively. The terms $X_{e,H}(z)$ and $X_{e,\mathrm{He}}(z)$ denote their ionization fractions. Since both are fully ionized at $z < 3$, we set $X_{e,H} = X_{e,\mathrm{He}} = 1$.

The host galaxy contribution, $\mathrm{DM}_{\mathrm{host}}$, depends on the local environment of the FRB and varies with redshift due to cosmological expansion. Although some studies assume a constant value (for example, $\mathrm{DM}_{\mathrm{host}} = 100~\mathrm{pc}\,\mathrm{cm}^{-3}$ \cite{tendulkar2017host}), this can yield unphysical results at low redshifts, including $\mathrm{DM}_{\mathrm{IGM}}$ negative values. To address this issue, we adopt the redshift-dependent parametrization proposed by \cite{zhang2020dispersion}, based on the IllustrisTNG simulation:

\begin{equation}\label{dmh}
    \mathrm{DM}_{\mathrm{host}} = A(1+z)^\alpha,
\end{equation}

\noindent where $A$ and $\alpha$ are free parameters that depend on the FRB type and host galaxy properties. For instance, repeating FRBs in spiral galaxies follow $\mathrm{DM}_{\mathrm{host}} = 34.72 (1+z)^{1.08}~\mathrm{pc}\,\mathrm{cm}^{-3}$, while non-repeating ones follow $\mathrm{DM}_{\mathrm{host}} = 32.97 (1+z)^{0.84}~\mathrm{pc}\,\mathrm{cm}^{-3}$.

The decomposition in Eq.~(\ref{dmobs}) isolates $\mathrm{DM}_{\mathrm{IGM}}$, which is crucial for using FRBs as cosmological probes, as discussed in the following sections.

\section{Distances in cosmology}\label{distances}
The comoving distance \(D_C\) is a key cosmological quantity, providing an invariant measure of spatial separation between an observer and a distant object, independent of cosmological expansion. 
In the  FLRW framework, \(D_C\) is defined as:

\begin{equation}\label{eq:comoving_distance}
D_C(z) = c \int_{0}^{z} \frac{dz'}{H(z')}.
\end{equation}
\noindent  A related scale is the Hubble distance (or Hubble radius) at the present epoch,
\begin{equation}
D_{H_0} = \frac{c}{H_0},
\end{equation}
with the general redshift-dependent definition
\begin{equation}
D_H(z) = \frac{c}{H(z)} .
\end{equation}
This scale is often used to normalize cosmological distances and should not be confused with causal horizons (particle or event horizons), which involve time integrals of $H(z)$.

On the other hand, the metric distance \(D_M\) is related to the comoving distance and depends on the curvature. For different values of the curvature parameter \(\Omega_k\), its functional form is \cite{weinberg1972gravitation}:

\begin{equation}\label{eq:DM}
D_M =
\begin{cases}
\frac{D_{H_0}}{\sqrt{\Omega_k}} \sinh\left(\sqrt{\Omega_k} \frac{D_C}{D_{H_0}} \right), & \text{if } \Omega_k > 0 , \\
D_C, & \text{if } \Omega_k = 0, \\
\frac{D_{H_0}}{\sqrt{|\Omega_k|}} \sin\left(\sqrt{|\Omega_k|} \frac{D_C}{D_{H_0}} \right), & \text{if } \Omega_k < 0 ,
\end{cases}
\end{equation}
where  $\Omega_k = 0$ for  a flat universe, $\Omega_k >0$  for an open universe and $\Omega_k <0$ for a closed universe.

In observational cosmology, absolute distance measurements are difficult to obtain; instead, dimensionless ratios  normalized by the sound horizon \( r_d \) are commonly used. Two ratios frequently used in BAO analysis are:

\begin{equation} \label{quantities}
\frac{D_M(z)}{r_d}, \quad \text{and} \quad \frac{D_H(z)}{r_d}.
\end{equation}
The sound horizon $r_d$, is the maximum distance that acoustic waves could travel in the primordial plasma before photon decoupling and is given by \cite{padilla2021cosmological}:

\begin{equation}
r_d = \int_{z_d}^{\infty} \frac{c_s dz}{H(z)},
\end{equation}

\noindent where \(c_s\) is the speed of sound in the primordial plasma and \(z_d\) denotes the baryon drag redshift, corresponding to the epoch when baryons decoupled from radiation. Measuring the scale \(r_d\) from BAO data provides an independent way to test cosmological models and infer the universe’s expansion rate.

The quantities  \( D_M / r_d \) and \( D_H / r_d \) serve as angular and radial standard rulers, respectively. \( D_M /r_d \) determines the angular scale of BAO features in galaxy distributions, while \( D_H/r_d \) determines the radial scale of the BAO peak in redshift-space distortions. BAO surveys measure the peak in both directions: perpendicular and along the line of sight, tracing \( D_M / r_d \) and \( D_H / r_d \), respectively. This dual measurement provides constraints on the expansion history $H(z)$ and the curvature parameter $\Omega_k$, offering an independent probe of the cosmological model.

In practice, the metric distance is not directly observable. Instead, we measure the angular diameter distance \( D_A \).
Due to cosmic expansion, \( D_A \) is related to the comoving transverse distance \( D_M \) by:

\begin{equation}\label{DAeq}
D_A(z) = \frac{D_M(z)}{1+z}.
\end{equation}





By combining distance ratios \( D_M/r_d \) and \( r_d/D_H \), we obtain an expression for \( D_A \) that depends only on observational quantities \cite{gong2024multiple,liu2025newest}:

\begin{equation}\label{BAODA}
D_A(z)=\frac{c}{(1+z)\,H(z)}\,\frac{D_M}{r_d}\,\frac{r_d}{D_H}.
\end{equation}

The equation above provides an angular diameter distance estimate that combines BAO ratios with an FRB-based reconstruction of the Hubble parameter. In this construction, the curvature parameter $\Omega_k$ does not enter explicitly, since the BAO measurements appear only through the ratio of their reported quantities and the sound horizon cancels. By contrast, the FRB-only approach reconstructs the comoving distance from the dispersion measure and then maps it into a transverse comoving distance using Eq.~(9), which explicitly involves $\Omega_k$. Thus, curvature dependence arises only in the FRB-only estimator, while the BAO+FRB estimator remains algebraically independent of $\Omega_k$.

 This formulation is useful for inferring \( D_A \) in a model-independent way, relying solely on measurements of the Hubble parameter and BAO-derived distance ratios. It enables cosmological tests without assuming a specific value of \( r_d \), making it robust to uncertainties in early-universe physics. We will use this equation later to combine FRBs and BAO da in section \ref{results}.

\section{Methodology}\label{sec4}
Following the discussions in the previous sections, we now describe the reconstruction of the comoving distance using real FRB data and its integration with BAO measurements. This section also outlines the technical procedures to derive angular diameter distances and constrain the curvature.

\subsection{Extracting $D_C$ and $D_A$ from $\langle \mathrm{DM_{IGM}}\rangle$}
To infer the comoving distance from the dispersion measure, we rewrite the Macquart relation, i.e., Eq.(\ref{DM_igm}), as:

\begin{equation} \label{digm2}
    \langle\mathrm{DM_{IGM}}(z)\rangle = A \int_0^z F(z')\, \dot D_C(z^\prime) dz',
\end{equation}
\noindent with 
\begin{equation}
A = \frac{3c \, \Omega_b H_0^2}{8 \pi G m_p} ,
\end{equation}
 where the free electron distribution in the IGM introduces an additional modulation  described by the function:
\begin{equation}
    F(z) = (1+z) f_{\mathrm{IGM}}(z) f_e(z),
\end{equation}
and  $\dot D_C(z)=dD_C(z)/dz$. 

The redshift-dependent factor $F(z)$ must be removed from  $\langle \mathrm{DM}_{\mathrm{IGM}}\rangle$. This is done by differentiating Eq.~(\ref{digm2}) with respect to redshift (note that Eq.~(\ref{eq:comoving_distance}) was also derived.)
\begin{eqnarray}\label{dotDc}
\dot{D_C}(z) =\frac{c}{H(z)} =\frac{1}{A F(z)} \frac{d\langle\mathrm{DM_{IGM}}(z)\rangle}{dz} \,,
\end{eqnarray}
and then integrating $c/H(z)$, thus yielding a direct  estimate of $D_C$ from the FRB data.


The key step in reconstructing $D_C$  is obtaining the derivative $d\langle\mathrm{DM_{IGM}(z)\rangle}\left(dz\right)^{-1}$. Although this method was first proposed in  \cite{yu2017measuring}, the lack of a sufficient number of localized FRBs at the time led to the use of simulations and Gaussian Processes. In contrast, here we apply the method to real data, using artificial neural networks (ANNs) to reconstruct the derivative and constrain cosmic curvature. In particular, we use the method developed in \cite{fortunato2025fast}, in which ANNs are trained to reconstruct the relation between \(\mathrm{DM_{IGM}}\) and \(z\) in a data-driven manner.

Specifically, we adopted a Multilayer Perceptron (MLP) architecture implemented using the \texttt{scikit-learn} library \cite{pedregosa2011scikit}, applying hyperparameter tuning via \texttt{GridSearchCV} to identify the optimal model. Various configurations of hidden layer sizes, activation functions, solvers, and learning rates were tested. The optimal model was selected based on cross-validation performance and learning curve analysis, ensuring that overfitting was minimized and that the network generalizes well to unseen data (see \cite{fortunato2025fast} for more details). Bootstrap resampling was used to estimate uncertainties in the ANNs predictions, enabling robust error propagation to the reconstructed derivative \(\frac{d\langle\mathrm{DM_{IGM}}\rangle}{dz}\) and, consequently, to \( D_C(z) \).

Once $D_C(z)$ is known,  the angular diameter distance can be derived solely from the FRB data using Eqs.~(\ref{eq:DM}) and~(\ref{DAeq}). From now on, we refer to the $D_A(z)$ obtained exclusively from FRB data   as \(D^{\mathrm{FRB}}_A(z) \). It is important to emphasize that the determination of both $D_C(z)$ and $D^{\mathrm{FRB}}_A(z)$  is entirely independent of any cosmological model and is derived directly from the data.
\subsection{Estimation of the cosmic curvature}
The key aspect of our approach to constraining spatial curvature is that FRBs play a dual role in the process. First, as described above, they allow us to directly obtain $D^{\mathrm{FRB}}_A(z)$  from the data.  Second, as demonstrated in \cite{fortunato2025fast}, the quantity  $\left(\frac{d\langle\mathrm{DM_{IGM}}(z)\rangle}{dz}\right)^{-1}$  enables the reconstruction of $H(z)$ and $H_0$ --- see Eq.~(\ref{dotDc}). The Hubble parameter derived from FRBs can also be combined with BAO measurements of the ratios $D_M/r_d$ and $D_H/r_d$, enabling the determination of the angular diameter distance via Eq.~(\ref{BAODA}) in a cosmology-independent manner, using only the data. We shall refer to the angular diameter distance obtained in this way as \( D^{\mathrm{BAO+FRB}}_A(z) \).    

 This approach allows us to compare  \( D^{\mathrm{FRB}}_A(z) \) and \( D^{\mathrm{BAO+FRB}}_A(z) \) to estimate the spatial curvature parameter \(\Omega_k\) in a data-driven manner.  In both cases of $D_A$ estimation, the Hubble constant $H_0$ is not externally imposed but is self-consistently derived by extrapolating the FRB-based reconstruction of $H(z)$ to redshift $z = 0$. This internal determination of $H_0$ helps mitigate the degeneracy between \(H_0\) and \(\Omega_k\), avoiding the need for external priors that could bias their coupling. Moreover, since  $H_0$ enters both $D_A$ and $D_C$ symmetrically, variations in its value do not affect the determination of $\Omega_k$ within this framework.

To constrain $\Omega_k$, we implement two distinct approaches: 

\textbf{Covariance matrix approach:} In our first approach, we account for the statistical correlation among the distance data points, primarily because both the angular diameter distance $D_A(z)$ and the comoving distance $D_C(z)$ depend on the same reconstructed Hubble parameter $H(z)$. To properly account for these correlations in the likelihood analysis, we construct a covariance matrix that incorporates both statistical uncertainties and correlated terms, following the strategy found in \cite{liu2025newest}, based on the error propagation theory \cite{owen2000statistical}. The likelihood function is given by
\begin{equation}
    \ln \mathcal{L} = -\frac{1}{2} \Delta D_i^T \, \mathbf{Cov}^{-1}_{ij} \, \Delta D_j,
\end{equation}
where the residuals are defined as $\Delta D = D_A^{\mathrm{BAO+FRB}}(z_i) - D_A^{\mathrm{FRB}}(z_i; \Omega_k)$, and the total covariance matrix is expressed as
\begin{equation}\label{coveq}
    \mathbf{Cov}_{ij} = \mathbf{Cov}^{\mathrm{stat}}_{ii} + \mathbf{Cov}^{\mathrm{corr}}_{ij}.
\end{equation}

The DESI DR2 survey \cite{karim2025desi} provides the covariance of the primary BAO observables
$\mathbf x\equiv\big(D_M/r_d,\;D_H/r_d\big)$ at each redshift, with a full
(off–diagonal) covariance matrix $\mathbf C_x$.
Since our analysis uses Eq. (\ref{BAODA}),
we propagate BAO errors to $D_A$ via the Jacobian \cite{owen2000statistical}
$\mathbf J_i=\big(\partial f/\partial(D_M/r_d),\;\partial f/\partial(D_H/r_d)\big)\big|_{z_i}$:
\begin{equation}
\mathbf C_{D_A}^{\mathrm{(BAO)}}=\mathbf J\,\mathbf C_x\,\mathbf J^{\!\top}.
\end{equation}
Following Eq.~(\ref{coveq}), only its diagonal enters $\mathbf{Cov}^{\mathrm{stat}}$,
and we add in quadrature the FRB-inferred $H(z)$ contribution on the data side:
\begin{equation}
\big(\mathbf{Cov}^{\mathrm{stat}}\big)_{ii}
=\big(\mathbf C_{D_A}^{\mathrm{(BAO)}}\big)_{ii}
+\left(\frac{\partial D_A}{\partial H}\sigma_H\right)^2_{z_i}.
\end{equation}

The correlated term captures the intrinsic coupling between
$D_A^{\mathrm{FRB}}$ and $D_C^{\mathrm{FRB}}$ induced by the same FRB–reconstructed
$H(z)$. We estimate it empirically from the FRB ensembles (bootstrap
realizations) as a cross–covariance:
\begin{eqnarray}
\mathbf{Cov}^{\mathrm{corr}}_{ij}
&=&\mathrm{Cov}\!\left[D_A^{\mathrm{FRB}}(z_i),\,D_C^{\mathrm{FRB}}(z_j)\right]\\
&=&\mathbb{E}\!\left[\big(D_A^{\mathrm{FRB}}-\mu_A\big)_i\,
\big(D_C^{\mathrm{FRB}}-\mu_C\big)_j\right].
\end{eqnarray}

Although the DESI collaboration provides a fiducial covariance matrix, it is derived under specific model assumptions and tailored to their reconstruction method. In contrast, our analysis relies on a model-independent reconstruction of \(H(z)\) and therefore requires a covariance matrix that accurately reflects our methodology. By constructing the covariance matrix directly from our reconstructed distances, we ensure internal consistency and avoid potential mismatches. As a cross-check, we repeated the inference using the DESI-provided covariance matrix (projected in the same way) and obtained posteriors for \(\Omega_k\) that were statistically indistinguishable from our baseline -- central values and uncertainties changed negligibly.

\textbf{Gaussian $\chi^2$ approach:} For comparison, we also implement a simpler approach in which all data points are treated as independent and Gaussian-distributed. In this case, we use the traditional $\chi^2$ minimization method:
\begin{equation}
    \chi^2 = \sum_i \frac{\left[D_A^{\mathrm{BAO+FRB}}(z_i) - D_A^{\mathrm{FRB}}(z_i; \Omega_k)\right]^2}{\sigma_{i}^2},
\end{equation}
where $\sigma_i$ represents the combined error from the reconstruction of $D_A$ and $D_C$. This method does not account for correlations among the data points and may underestimate the uncertainty, but it serves as a useful baseline.
The likelihood therefore compares a curvature--independent quantity, defined in Eq.~(13), with a curvature--dependent one obtained via Eq.~(9), and the sensitivity to $\Omega_k$ emerges solely from this comparison.

In summary, the \textsc{COV} analysis uses the full matrix $\mathbf{Cov}=\mathrm{diag}\!\big(\mathbf{Cov}^{\mathrm{stat}}\big)+\mathbf{Cov}^{\mathrm{corr}}$,
where the BAO contribution to $D_A$ is rigorously propagated from DESI’s (off–diagonal) covariance via the Jacobian, while the correlated FRB term is measured from the FRB ensemble. For comparison, the Gaussian case adopts only the diagonal variance
$\sigma_i^2=\big(\mathbf C_{D_A}^{\mathrm{(BAO)}}\big)_{ii}
+\left(\partial D_A/\partial H\,\sigma_H\right)^2_{z_i}$,
i.e.\ it neglects $\mathbf{Cov}^{\mathrm{corr}}$.






\section{Data}\label{data}
To implement our analysis, we use a subset of 120 from the 131 localized FRBs presented in table \ref{tab:frb_sample}. In addition to requiring known redshifts and dispersion measures, we impose a selection criterion also introduced in \cite{wang2025probing} to ensure the physical consistency of the intergalactic medium contribution to the dispersion measure, \( \mathrm{DM}_{\mathrm{IGM}} \). Specifically, we require
\[
\mathrm{DM}_{\mathrm{obs}} - \mathrm{DM}_{\mathrm{MW}} > 80~\mathrm{pc~cm^{-3}},
\]
which accounts for a typical range of halo contributions, \( \mathrm{DM}_{\mathrm{halo}} \sim 50\text{--}80~\mathrm{pc~cm^{-3}} \). This criterion serves as a practical proxy to enforce \( \mathrm{DM}_{\mathrm{obs}} - \mathrm{DM}_{\mathrm{MW}} - \mathrm{DM}_{\mathrm{halo}} > 0 \), ensuring that the inferred \( \mathrm{DM}_{\mathrm{IGM}} \) remains physically meaningful and avoids negative values. This cut also helps to eliminate FRBs with low total extragalactic dispersion, which could lead to biased inferences of the Hubble parameter, particularly by favouring artificially high values of \( H_0 \) within a narrow prior range, as shown in \cite{wang2025probing}. Furthermore, this selection indirectly filters out several low-redshift FRBs that may be heavily influenced by peculiar velocities of their host galaxies and the local cosmic web. By excluding these events, we minimize the impact of non-cosmological effects on the reconstruction of \( H(z) \) and improve the robustness of the derived  distances. 

\begin{table*}[t]
    \centering
    \begin{tabular}{l c c c c}
        \hline
        Tracer & $z$ & $D_M/r_d$ & $D_H/r_d$ & Ref. \\
        \hline
        LRG1 & 0.510 & $13.588\pm0.167$ & $21.863\pm0.425$ & \cite{karim2025desi} \\
        LRG2 & 0.706 & $17.351\pm0.177 $ & $19.455\pm0.330$ & \cite{karim2025desi} \\
        LRG3+ELG1& 0.934 & $21.576\pm0.152$ & $17.641 \pm 0.193$ & \cite{karim2025desi} \\
        ELG2 & 1.321 & $27.601\pm0.318$ & $14.176\pm0.221$ & \cite{karim2025desi} \\
        QSO & 1.484 & $30.512\pm0.760$ & $12.817\pm0.516$ & \cite{karim2025desi} \\
        Lya & 2.330 & $338.988\pm0.531$ & $8.632\pm0.101$ & \cite{karim2025desi} \\
        
        \hline
    \end{tabular}
    \caption{Measurements of $D_M/r_d$ and $D_H/r_d$ from DESI DR2 survey.}
    \label{tab:BAO_measurements}
\end{table*}

In addition, seven events were excluded from our analysis for specific reasons. First, FRB200110E is omitted because it is located at an extremely close distance of only $3.6~\mathrm{Mpc}$, within a globular cluster in the M81 galaxy \cite{kirsten2022repeating, bhardwaj2021nearby}. Its proximity implies a negligible contribution from the intergalactic medium, making it unsuitable for cosmological inference. Second, FRB210117 and FRB190520 display dispersion measures significantly exceeding theoretical expectations based on their redshift. These anomalies classify them as statistical outliers, probably influenced by a dense or highly magnetized local environment in their host galaxy \cite{bhandari2023nonrepeating, gordon2023demographics,niu2022repeating}. \noindent A practical limitation of our reconstruction is the sparse sampling at high redshift (\(z \gtrsim 1\)). Our baseline analysis includes all available localized FRBs after cleaning the dataset (120 events in total), among which only a handful lie near or above \(z \sim 1\) (notably FRB220610, FRB221029, FRB230521, FRB240123, and the most distant FRB240304B at \(z=2.148\)). In this regime, the ANN is weakly constrained by the data, which naturally broadens the predictive variance and, consequently, inflates the propagated uncertainties in \(H(z)\), \(D_C(z)\), and \(D_A^{\mathrm{FRB}}(z)\). This effect is expected and reflects limited leverage rather than a modeling bias. Since our curvature inference is most strongly informed by the redshift range where BAO measurements exist (\(z \lesssim 2.5\)), the net impact is a conservative widening of error bands at high \(z\) without qualitatively altering the central trends used to constrain \(\Omega_k\).


For the analysis that follows, we employ a total of ten BAO data points in the range of the reconstructed FRBs redshifts, z = [0, 1.5]: five from the BOSS/eBOSS programs and five from the recent DESI DR2 release,  summarized in Table~\ref{tab:BAO_measurements}. These measurements are used in combination with our FRB and ANN-based reconstructions of $H(z)$ and distances, to constrain the cosmic curvature parameter $\Omega_k$.

\section{Results and Discussions}\label{results}
\begin{figure*}[htbp]
  \centering
  \includegraphics[width=.49\textwidth]{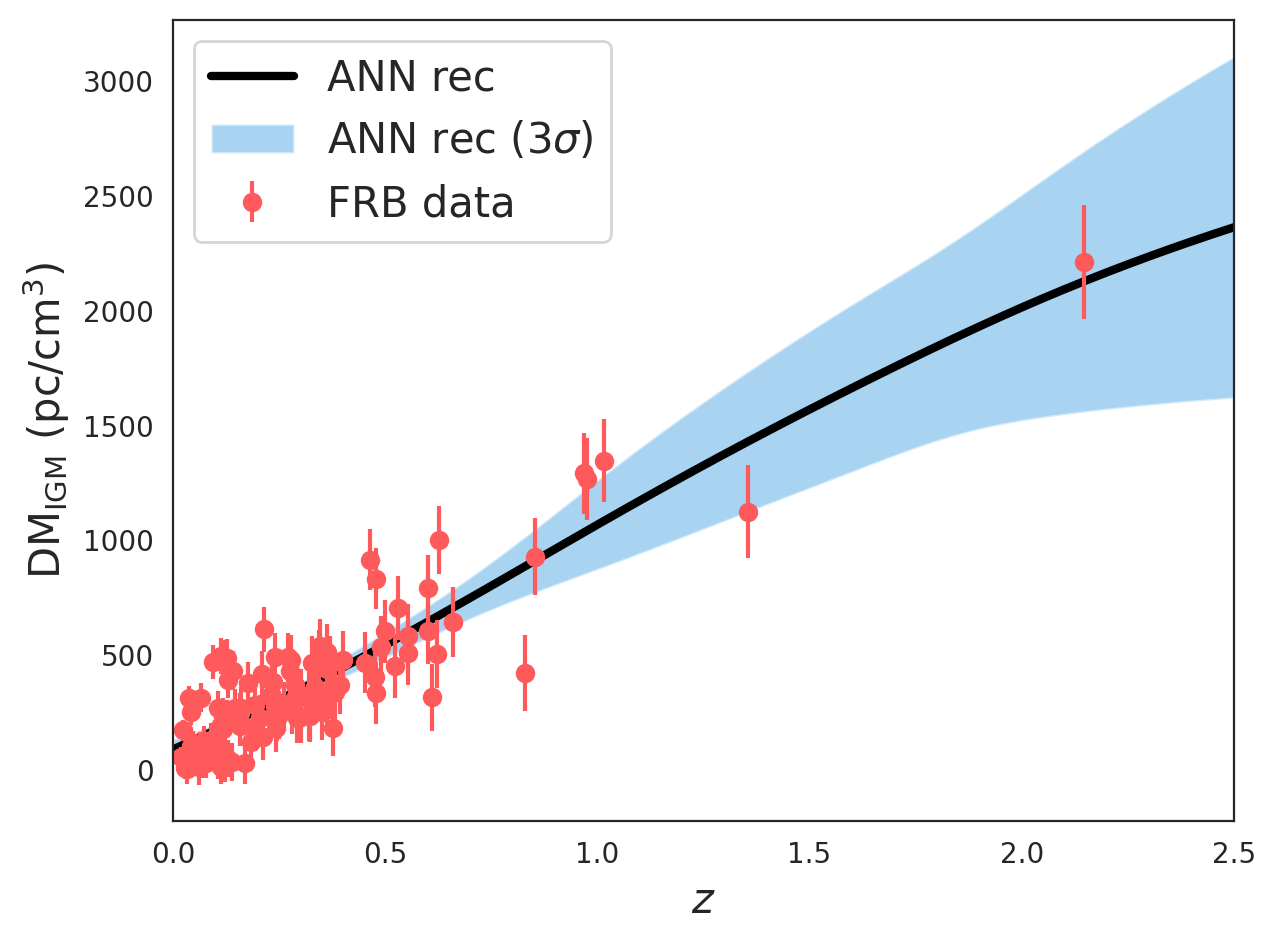}
  \includegraphics[width=.49\textwidth]{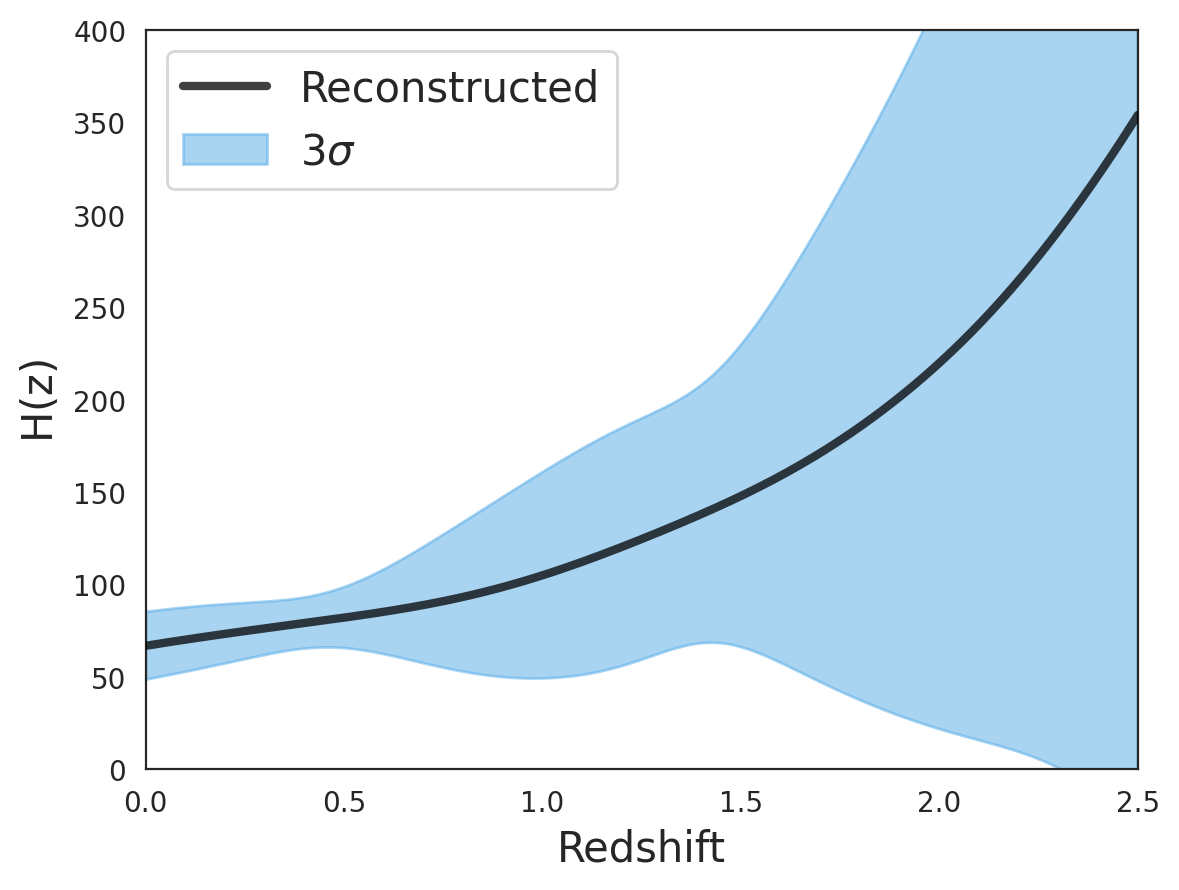}

  \caption{Reconstructed FRBs observables. Left: redshift evolution of the average IGM dispersion measure $\langle \mathrm{DM}_{\mathrm{IGM}}(z) \rangle$. Right: Hubble parameter $H(z)$ inferred from FRBs.}
  \label{fig:dp_reconstruction}
\end{figure*}

To apply the methodology described above, we first subtract the contributions from the Milky Way, the halo, and the host galaxy in order to obtain $\mathrm{DM_{IGM}}$ from the observed dispersion measures of the subset of 120 FRBs described in Sec.~\ref{data}. 
Then we reconstruct the comoving distance \( D_C(z) \). The first step involves determining the redshift evolution of the average intergalactic dispersion measure, \(\langle \mathrm{DM}_{\mathrm{IGM}}(z) \rangle\). As detailed in Sec.~\ref{sec4}, this is performed using an ANN algorithm (see \cite{fortunato2025fast} for more details). The result is shown in panel (a) in Figure~\ref{fig:dp_reconstruction}.  The red points with error bars represent the individual FRB measurements and their associated uncertainties, while the solid black curve corresponds to the ANN-based reconstruction. The shaded region around it denotes the \(3\sigma\) confidence interval. 

\begin{figure}[htbp]
  \includegraphics[width=.49\textwidth]{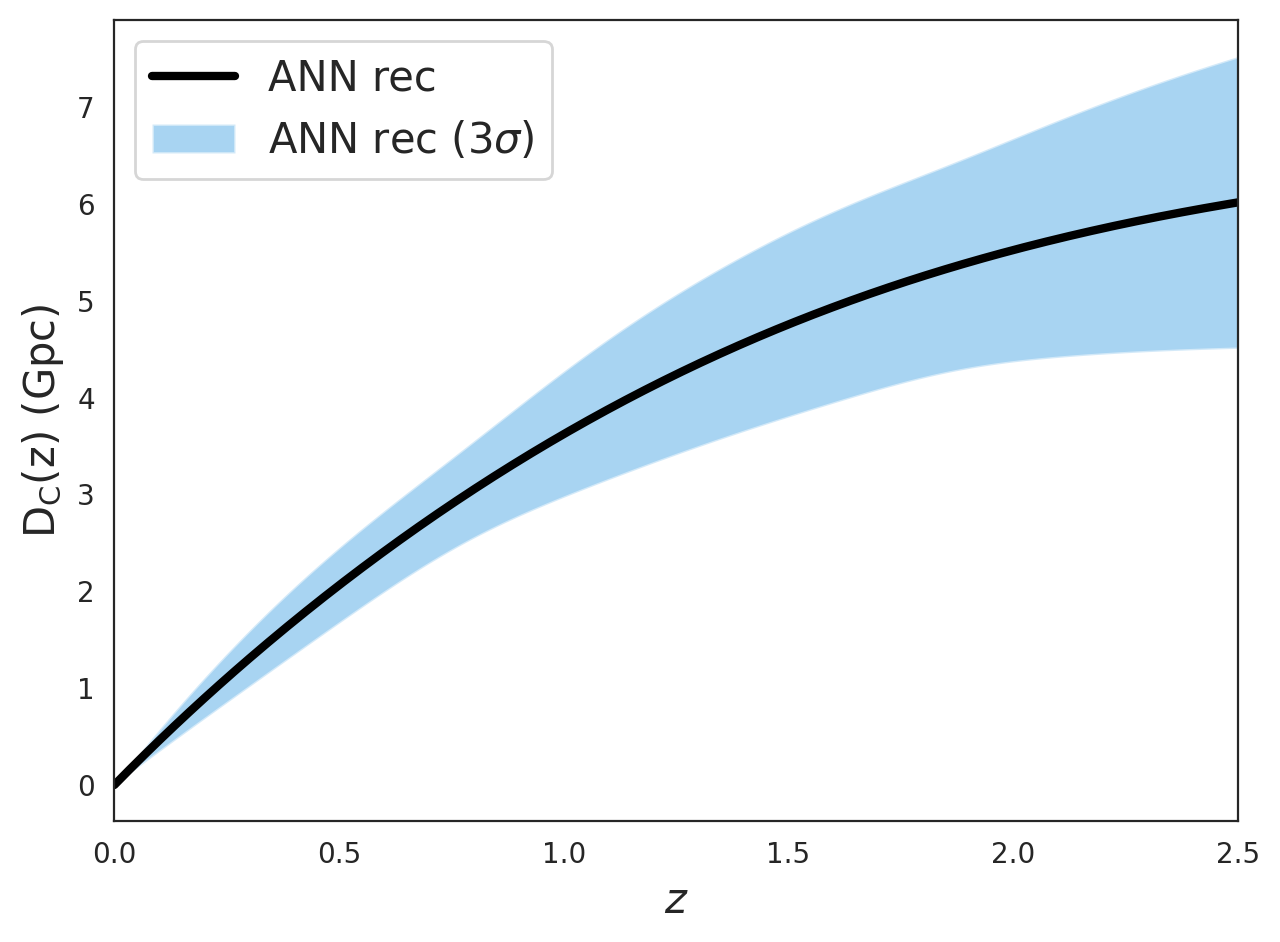}

  \caption{Comoving distance $D_C(z)$ reconstructed from FRBs.}
  \label{fig:dc_reconstruction2}
\end{figure}

Once \(\langle \mathrm{DM}_{\mathrm{IGM}}(z) \rangle\) is reconstructed, it is used to derive the Hubble parameter $H(z)$ and by extrapolating the reconstruction to $z=0$ we obtain
$H_0 = 68.74 \pm 6.11\ \mathrm{km\,s^{-1}\,Mpc^{-1}}$. Panel (b) of
Figure \ref{fig:dp_reconstruction} shows the reconstructed $H(z)$ curve with $3\sigma$ bands. This value is then applied throughout our methodology as described in Section \ref{sec4} to recover the comoving distance \( D_C(z) \) (see Eq.~(\ref{dotDc})). The resulting comoving distance curve is displayed in Figure~\ref{fig:dc_reconstruction2}. The dashed black line shows the reconstructed \( D_C(z) \), while the shaded blue band represents the associated \(3\sigma\) uncertainty. This reconstruction provides a continuous cosmological model-independent estimate of the comoving distance across a range of redshifts,  from low redshifts up to \( z \sim 2.5 \), where BAO measurements are also available for comparison. Subsequently, \(D^{\mathrm{FRB}}_A(z) \) and  \(D^{\mathrm{BAO+FRB}}_A(z)  \) are computed.

\begin{figure*}[htbp]
  \centering
  \includegraphics[width=.49\textwidth]{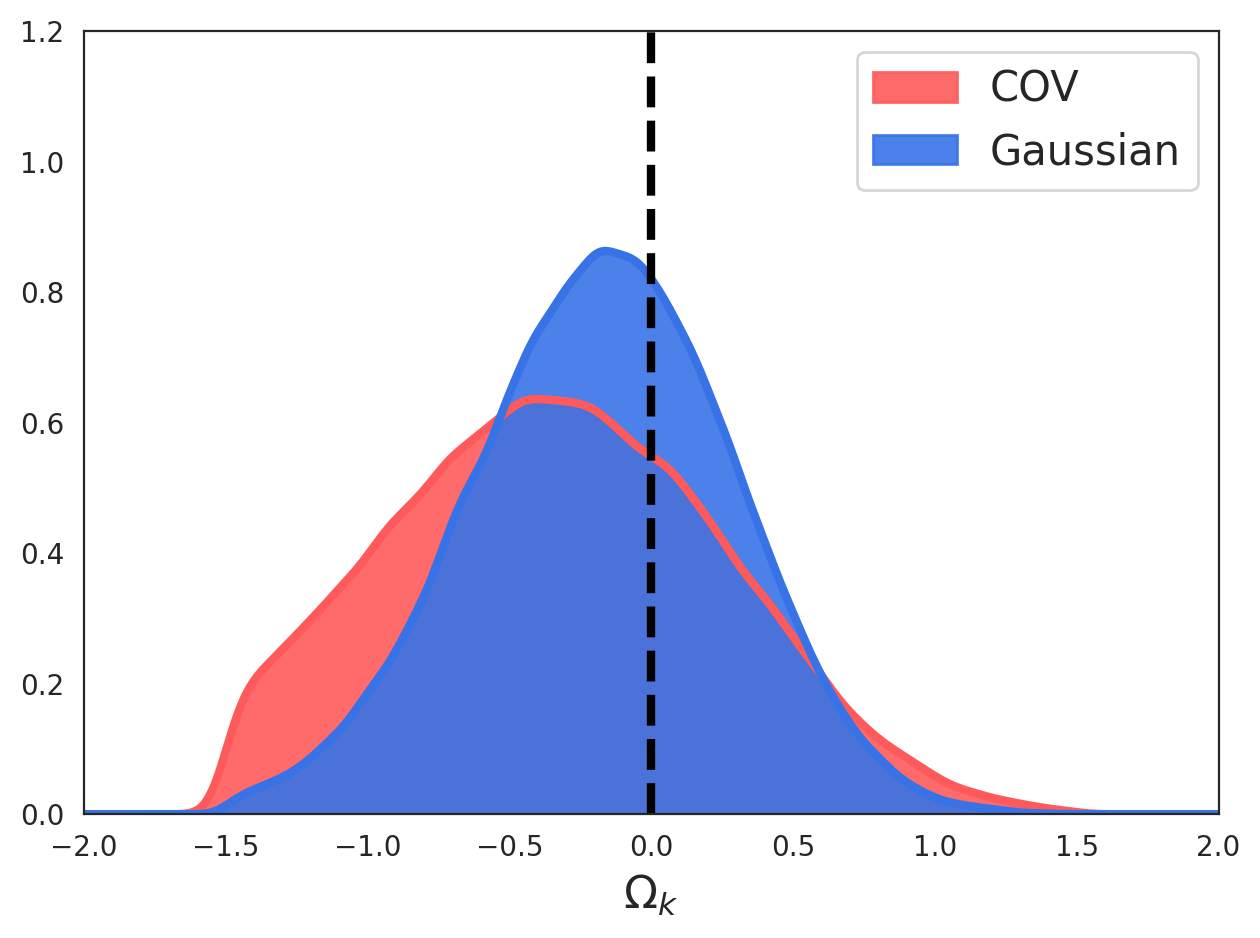}
  \includegraphics[width=.49\textwidth]{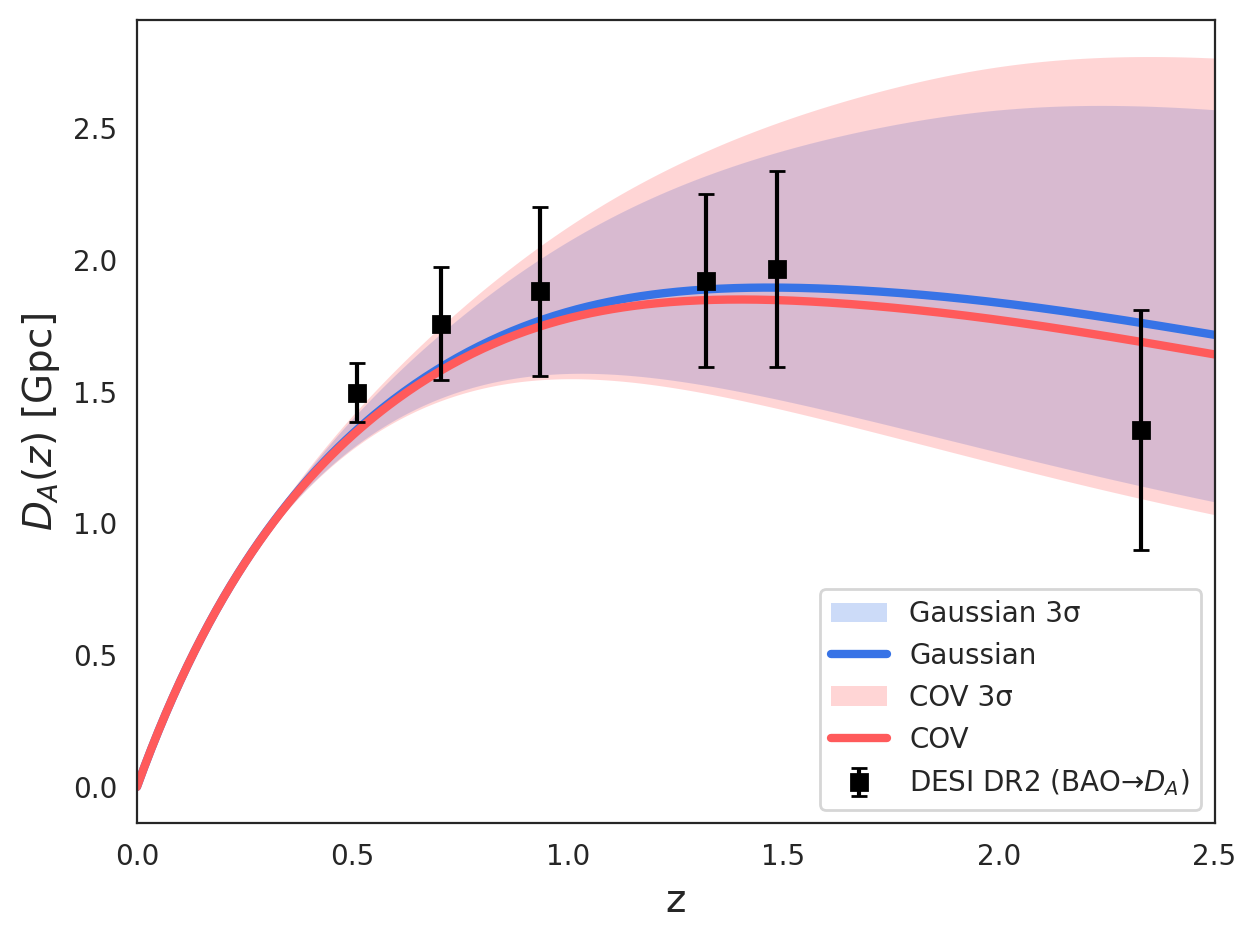}
    \caption{Left: One–dimensional posterior for the spatial curvature parameter $\Omega_k$ obtained with our two likelihoods. 
The red filled curve (COV) uses the full covariance between the FRB–derived $D_C$ and $D_A$ together with the BAO error propagation;  the blue filled curve (Gaussian) adopts a diagonal treatment. The vertical dashed line marks $\Omega_k=0$. Right: Black points with $3\sigma$ bars are $D_A^{\rm BAO+FRB}(z)$ from Eq.~\ref{BAODA}, using DESI~DR2 data with the FRB–based $H(z)$ and propagated BAO uncertainties. Colored curves show the FRB–only estimator $D_A^{\rm FRB}(z;\Omega_k)$ evaluated at the mean $\Omega_k$ of each analysis; the shaded bands denote credible regions.}
\label{fig:corner_plots}
\end{figure*}

To constrain $\Omega_k$, we explore the likelihood using the \texttt{emcee} Python module, a Markov Chain Monte Carlo (MCMC) ensemble sampler~\cite{foreman2013emcee}. The prior for the Hubble constant is taken as the extrapolated value $H_0 = H(z = 0)$,  obtained from the ANN reconstruction of the FRB data. The final results from both approaches (covariance matrix and Gaussian $\chi^2$ minimization) are presented in Figure~\ref{fig:corner_plots}, panel (a). These results demonstrate that the combined use of FRBs and BAO data provides a robust and independent constraint on the spatial geometry of the Universe.

From the covariance-based analysis by employing our 120 FRBs and the DESI DR2 datasets, we derive an estimated $\Omega_k=-0.31 \pm 0.57$ for the cosmic curvature parameter. In contrast, neglecting the off-diagonal elements of the covariance matrix yields a bit tighter posterior distribution, $\Omega_k=-0.13 \pm 0.46$. Despite variations in central values, all estimates remain consistent with a spatially flat Universe ($\Omega_k = 0$) at the $1\sigma$ level. The integration of FRB-based comoving distance reconstruction serves as an independent and complementary cosmological probe, reinforcing the growing utility of FRBs in precision cosmology.

In panel (b) of Figure~\ref{fig:corner_plots} we compare the FRB-only reconstruction of the angular–diameter distance, $D_A^{\rm FRB}(z;\Omega_k)$, obtained with our two likelihood treatments against the discrete $D_A^{\rm BAO+FRB}$ estimates at the DESI DR2 effective redshifts -- black squares. Solid curves show the posterior–mean $D_A^{\rm FRB}(z)$ for each treatment, while the shaded bands indicate the propagated $3\sigma$ uncertainty, which naturally widens toward high $z$ where the FRB sample is sparse. The DESI points (with $1\sigma$ error bars) are derived from the measured BAO ratios via Eq.~\ref{BAODA} combined with the FRB-based $H(z)$ reconstruction. The two FRB-only solutions agree closely over $0.4\!\lesssim\!z\!\lesssim\!2$, with the COV band being broader—as expected when correlations are retained—yet both envelopes encompass the BAO+FRB determinations within uncertainties. The peak around $z\!\sim\!1.6$ is consistent with the expected maximum of $D_A(z)$ in late-time cosmologies, and the overall consistency of the three estimators supports our curvature analysis.

Taken together, these results demonstrate that ANN reconstructions of comoving distances from FRBs, when combined with state-of-the-art BAO measurements, provide competitive constraints on spatial curvature, comparable to those obtained from other late-time cosmological probes \cite{liu2025newest, gong2024multiple, mukherjee2022constraining, wang2021machine}. In our analysis, both likelihood treatments are statistically consistent with spatial flatness at the $1\sigma$ level, but they behave differently with respect to uncertainty budgeting. The Gaussian approximation produces visibly tighter credible intervals, yet it risks understating the true errors because it neglects cross–terms induced by the common FRB–based $H(z)$ entering both $D_A$ and $D_C$. When we retain the full covariance, the posteriors broaden as expected, yielding a more conservative and robust assessment of parameter uncertainties. Importantly, the mild preference for $\Omega_k<0$ persists in both treatments, and the shift of the posterior mean when toggling the covariance is small, indicating that our qualitative conclusion is not driven by the covariance choice.

 \begin{figure*}
    \centering
     \includegraphics[width=.7\textwidth]{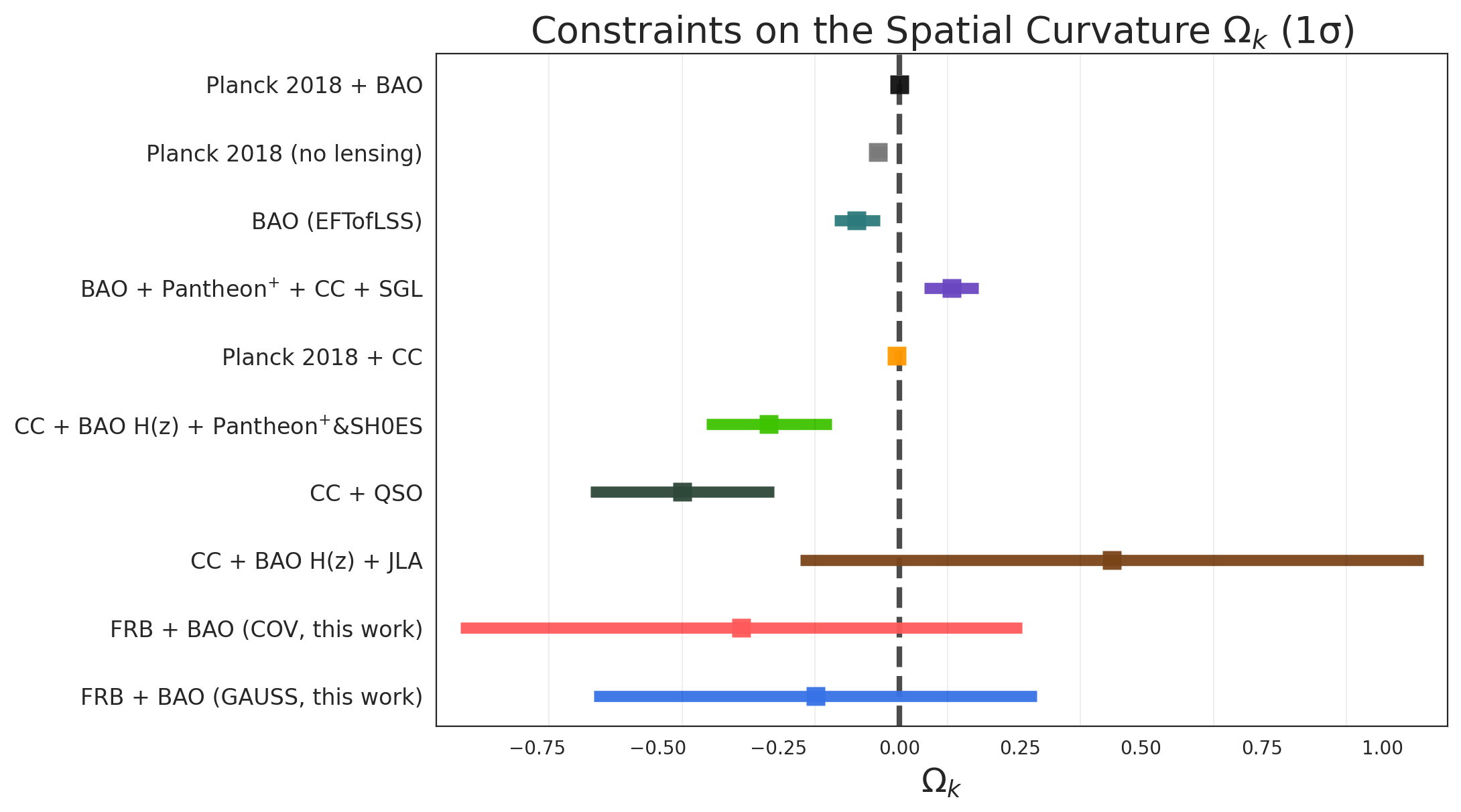}
    \caption{%
    Constraints on the spatial curvature \(\Omega_k\).
    Horizontal bars show \(1\sigma\) confidence intervals for our
    FRB + BAO (COV) and FRB + BAO (GAUSS) analyses, alongside a set of literature determinations
    (Planck 2018 + BAO; Planck 2018 (no lensing); EFTofLSS/BAO; BAO + Pantheon+; CC + SGL; Planck 2018 + CC; etc.).
    The vertical dashed line marks \(\Omega_k=0\).
    Incorporating the full covariance (COV) yields broader, more conservative constraints than the Gaussian approximation (GAUSS), yet both remain consistent with spatial flatness at \(1\sigma\).
    }
    \label{fig:ok_comparison}
\end{figure*}

Figure~\ref{fig:ok_comparison} summarizes our \(\Omega_k\) posteriors for FRB+BAO (COV) and FRB+BAO (Gaussian) together with a representative selection of external results. Error bars preserve the original symmetry/asymmetry reported by each source. Placed in the broader context of the literature, our result aligns with determinations that favor near–flat geometries when large–scale correlations and/or late–time anchors are incorporated (e.g., Planck 2018 + BAO~\cite{aghanim2020planck}; Planck 2018 + CC~\cite{vagnozzi2021eppur}), while remaining compatible with analyses that report a mild preference for closed spatial sections under more restrictive data choices or modeling assumptions (e.g., Planck 2018 without lensing~\cite{di2020planck}, EFTofLSS/BAO~\cite{glanville2022full}). For late–Universe, non–CMB comparisons we include: CC + BAO \(H(z)\) + JLA~\cite{wang2017model}; CC + BAO \(H(z)\) + Pantheon\(^+\)\,\&\,SH0ES~\cite{qi2023model}; CC + QSO (X–ray/UV quasars)~\cite{dinda2023constraints}; and a recent joint late–time analysis BAO + SNe (DES) + CC + strong–lensing time delays~\cite{wu2025measuring}. We emphasize that, unlike many CMB–anchored constraints, our inference exploits a model–independent, late–time reconstruction of \(H(z)\) from FRBs confronted with BAO ratios, helping to mitigate curvature–\(H_0\) degeneracies and providing an independent cross–check of spatial geometry.

\section{Conclusions}\label{conclu}
In this work, we presented a fully data-driven 
strategy to constrain the spatial curvature of the Universe by leveraging the synergy between FRBs and BAO measurements. Reconstructing the Hubble parameter $H(z)$ using an ANN trained on a set of localized FRBs, we derived the comoving distance $D_C(z)$ and subsequently estimated $D_A(z)$ to compare with the same quantity obtained from BAO+FRB data. This dual-path approach enabled robust constraints on the curvature parameter $\Omega_k$, independent of early-universe assumptions or fiducial cosmological models. Our analysis yields $\Omega_k=-0.31 \pm 0.57$ (full covariance)
and $\Omega_k=-0.13 \pm 0.46$ (Gaussian approximation).
While the uncorrelated treatment produces tighter intervals, it risks underestimating uncertainties,
we therefore adopt the covariance-based constraint as our headline result.

Our results indicate that, within current observational limits, all estimates remain consistent with a spatially flat Universe. However, a persistent mild preference for negative curvature is observed across multiple statistical treatments, suggesting that future improvements in data quantity and quality could be decisive. Incorporating the full covariance matrix in the likelihood analysis yields more conservative yet reliable uncertainty estimates, underscoring the importance of properly accounting for statistical correlations in precision cosmology.

Looking ahead, enhancing the constraining power of FRB-based curvature measurements will require advances on both observational and methodological fronts. Increasing the number of localized FRBs with secure redshift determinations will reduce statistical uncertainties and improve redshift coverage.  The number of FRBs detected is indeed rapidly increasing with the advent of large-scale radio surveys and real-time localization capabilities, such as the  Canadian Hydrogen Intensity Mapping Experiment (CHIME) \cite{amiri2022overview}, the Australian Square Kilometre Array Pathfinder (ASKAP) \cite{qiu2023systematic}, the improved version of the Karoo Array Telescope (MeerKAT) \cite{tian2024detection}. At the same time, improved modeling of host galaxy and halo contributions to the dispersion measure will mitigate systematic errors in the extraction of $\mathrm{DM}_{\mathrm{IGM}}$. On the analytical side, incorporating hybrid inference techniques that combine neural networks with advanced statistical frameworks may enhance the reconstruction accuracy and stability.

As new FRBs discoveries expand the redshift baseline and improve the resolution of the available datasets, their integration with BAO measurements promises to yield increasingly precise and independent insights into the geometry of the cosmos. This work reinforces the case for FRBs as a valuable cosmological probe and sets the stage for their broader inclusion in future large-scale structure analyses.

\acknowledgments

JASF thanks FAPES and the National Research Foundation (NRF) for their financial support. WSHR acknowledges partial financial support from FAPES. G.E.R. acknowledges financial support from the State Agency for Research of the Spanish Ministry of Science and Innovation under grant PID2022-136828NB-C41/AEI/10.13039/501100011033/, and by
“ERDF A way of making Europe”, by the “European Union”. He also thanks the support from PIP 0554 (CONICET). JASF thanks Amanda Weltman and Surajit Kalita for fruitful discussions.

\appendix
\subsection{Dataset}

In this appendix we present the Full Localized FRBs sample, including the ones used in this work and those we neglected for the reasons described in the main text. Reported observables are the measured dispersion measure $\mathrm{DM}_{\rm obs}$ and the Milky Way contribution from the NE2001 model.


\onecolumngrid
\begingroup
\setlength\tabcolsep{5pt}
\renewcommand{\arraystretch}{1.05}

\begin{longtable}{|lcccc|}
\caption{ Localized FRBs}
\label{tab:frb_sample}\\
\toprule
\hline
FRB & $z$ & $\mathrm{DM}_{\rm obs}$ & $\mathrm{DM}_{\rm MW}$ (NE2001) & Ref. \\
\midrule
\endfirsthead

\bottomrule
\endlastfoot

\hline
FRB200120E & 0.0008   & 87.8 & 40.8 & \cite{connor2024gas} \\
FRB181030A & 0.00385  & 103.396 & 41.1 & \cite{lanman2022sudden} \\
FRB171020A & 0.008672 & 114.1  & 38.0 & \cite{mahony2018search} \\
FRB220319D & 0.011228 & 110.98 & 65.25 & \cite{ravi2023deep} \\
FRB231129A & 0.019000 & 198.50 & 58.3 & \cite{amiri2025catalog} \\
FRB240210A & 0.023686 & 283.73 & 28.7 & \cite{shannon2025commensal} \\
FRB181220A & 0.027460 & 208.66 & 118.5 & \cite{bhardwaj2024host} \\
FRB231230A & 0.029800 & 131.40 & 23.0 & \cite{amiri2025catalog} \\
FRB181223C & 0.030240 & 111.61 & 19.9 & \cite{bhardwaj2024host} \\
FRB190425A & 0.03122 & 127.78  & 48.7 & \cite{bhardwaj2024host} \\
FRB180916B & 0.033700 & 349.35 & 199.0 & \cite{marcote2020repeating} \\
FRB230718A & 0.0357 & 477.0 & 420.6 & \cite{glowacki2024h} \\
FRB231120A & 0.036800 & 437.74 & 43.8 & \cite{sharma2024preferential} \\
FRB240201A & 0.042729 & 374.50 & 38.6 & \cite{shannon2025commensal} \\
FRB220207C & 0.043040 & 262.38 & 76.1 & \cite{law2024deep} \\
FRB211127I & 0.046900 & 234.83 & 42.5 & \cite{glowacki2023wallaby} \\
FRB201123A & 0.050700 & 433.55 & 251.7 & \cite{rajwade2022first} \\
FRB230926A & 0.055300 & 222.80 & 52.6 & \cite{amiri2025catalog} \\
FRB200223B & 0.060240 & 201.80 & 45.6 & \cite{ibik2024proposed} \\
FRB190303A & 0.064000 & 223.20 & 29.8 & \cite{michilli2023subarcminute} \\
FRB231204A & 0.064400 & 221.00 & 34.9 & \cite{amiri2025catalog} \\
FRB231206A & 0.065900 & 457.70 & 59.1 & \cite{amiri2025catalog} \\
FRB210405I & 0.066 & 565.17 & 516.1 & \cite{driessen2024frb}\\
FRB180814A & 0.068 & 190.9 & 87.5 & \cite{michilli2023subarcminute} \\
FRB211212A & 0.070700 & 206.00 & 38.8 & \cite{Deller2025} \\
FRB231005A & 0.071300 & 189.40 & 33.5 & \cite{amiri2025catalog} \\
FRB190418A & 0.071320 & 182.78 & 70.2 & \cite{bhardwaj2024host} \\
FRB231123A & 0.072900 & 302.10 & 89.7 & \cite{amiri2025catalog} \\
FRB220912A & 0.0771 & 219.46 & 125.2 & \cite{ravi2023deep} \\
FRB231011A & 0.078300 & 186.30 & 70.4 & \cite{amiri2025catalog} \\
FRB220509G & 0.089400 & 269.53 & 55.6 & \cite{law2024deep} \\
FRB230124A & 0.093900 & 590.57 & 38.6 & \cite{sharma2024preferential} \\
FRB201124A & 0.098000 & 413.52 & 139.9 & \cite{ravi2022host} \\
FRB230708A & 0.105000 & 411.51 & 60.3 & \cite{shannon2025commensal} \\
FRB231223C & 0.105900 & 165.80 & 47.9 & \cite{amiri2025catalog} \\
FRB191106C & 0.107750 & 332.20 & 25.0 & \cite{ibik2024proposed} \\
FRB231128A & 0.107900 & 331.60 & 64.7 & \cite{amiri2025catalog} \\
FRB230222B & 0.110000 & 187.80 & 27.7 & \cite{amiri2025catalog} \\
FRB231201A & 0.111900 & 169.40 & 70.0 & \cite{amiri2025catalog} \\
FRB220914A & 0.113900 & 631.28 & 54.7 & \cite{law2024deep} \\
FRB190608B & 0.117780 & 338.70 & 37.3 & \cite{chittidi2021dissecting} \\
FRB230703A & 0.118400 & 291.30 & 26.9 & \cite{amiri2025catalog} \\
FRB240213A & 0.118500 & 357.40 & 40.0 & \cite{connor2024gas} \\
FRB230222A & 0.122300 & 706.10 & 134.2 & \cite{amiri2025catalog} \\
FRB190110C & 0.122440 & 221.60 & 37.1 & \cite{ibik2024proposed} \\
FRB230628A & 0.127000 & 344.95 & 39.0 & \cite{sharma2024preferential} \\
FRB240310A & 0.127000 & 601.80 & 30.1 & \cite{shannon2025commensal} \\
FRB210807D & 0.129300 & 251.90 & 121.2 & \cite{glowacki2023wallaby} \\
FRB240114A & 0.130000 & 527.65 & 49.7 & \cite{kumar2024varying} \\
FRB240209A & 0.138400 & 176.57 & 55.5 & \cite{shah2025repeating} \\
FRB210410D & 0.141500 & 571.20 & 56.2 & \cite{caleb2023subarcsec} \\
FRB230203A & 0.146400 & 420.10 & 67.3 & \cite{amiri2025catalog} \\
FRB231226A & 0.156900 & 329.90 & 38.1 & \cite{shannon2025commensal} \\
FRB230526A & 0.157000 & 361.40 & 31.9 & \cite{shannon2025commensal} \\
FRB220920A & 0.158239 & 314.99 & 39.9 & \cite{law2024deep} \\
FRB200430A & 0.160800 & 380.10 & 27.2 & \cite{heintz2020host} \\
FRB190203A & 0.170000 & 134.40 & 20.3 & \cite{tyul2025pushchino} \\
FRB210603A & 0.177200 & 500.15 & 39.5 & \cite{cassanelli2024fast} \\
FRB220529A & 0.183900 & 246.00 & 40.0 & \cite{li2025active} \\
FRB230311A & 0.191800 & 364.30 & 92.5 & \cite{amiri2025catalog} \\
FRB220725A & 0.192600 & 290.40 & 30.7 & \cite{shannon2025commensal} \\
FRB121102A & 0.192730 & 557.00 & 188.4 & \cite{tendulkar2017host} \\
FRB221106A & 0.204400 & 343.80 & 34.8 & \cite{shannon2025commensal} \\
FRB240215A & 0.210000 & 549.50 & 47.9 & \cite{connor2024gas} \\
FRB230730A & 0.211500 & 312.50 & 85.2 & \cite{amiri2025catalog} \\
FRB210117A & 0.214500 & 729.10 & 34.4 & \cite{bhandari2023nonrepeating} \\
FRB221027A & 0.229000 & 452.50 & 47.2 & \cite{sharma2024preferential} \\
FRB191001A & 0.234000 & 506.92 & 44.2 & \cite{heintz2020host} \\
FRB190714A & 0.236500 & 504.13 & 38.5 & \cite{heintz2020host} \\
FRB221101B & 0.239500 & 491.55 & 131.2 & \cite{sharma2024preferential} \\
FRB220825A & 0.241397 & 651.24 & 78.5 & \cite{law2024deep} \\
FRB190520B & 0.2418 & 1204.7 & 63 & \cite{ocker2022large} \\
FRB191228A & 0.243200 & 297.50 & 32.9 & \cite{bhandari2022characterizing} \\
FRB231017A & 0.245000 & 344.20 & 31.2 & \cite{amiri2025catalog} \\
FRB220307B & 0.248123 & 499.27 & 128.2 & \cite{law2024deep} \\
FRB221113A & 0.250500 & 411.03 & 91.7 & \cite{sharma2024preferential} \\
FRB220831A & 0.262 & 1146.25 & 126.8 & \cite{connor2024gas} \\ 
FRB231123B & 0.262100 & 396.86 & 40.3 & \cite{amiri2025catalog} \\
FRB230307A & 0.270600 & 608.85 & 37.6 & \cite{sharma2024preferential} \\
FRB221116A & 0.276400 & 643.45 & 132.3 & \cite{sharma2024preferential} \\
FRB220105A & 0.278500 & 583.00 & 22.0 & \cite{shannon2025commensal} \\
FRB210320C & 0.279700 & 384.80 & 39.3 & \cite{glowacki2023wallaby} \\
FRB221012A & 0.284669 & 441.08 & 54.3 & \cite{law2024deep} \\
FRB240229A & 0.287000 & 491.15 & 38.0 & \cite{connor2024gas} \\
FRB190102C & 0.291200 & 364.50 & 57.4 & \cite{macquart2020census} \\
FRB220506D & 0.300390 & 396.97 & 84.6 & \cite{law2024deep} \\
FRB230501A & 0.301500 & 532.47 & 125.7 & \cite{sharma2024preferential} \\
FRB230503E & 0.320000 & 483.74 & 88.0 & \cite{pastor2025localisation} \\
FRB180924B & 0.321200 & 361.42 & 40.5 & \cite{bannister2019single} \\
FRB231025B & 0.323800 & 368.70 & 48.6 & \cite{amiri2025catalog} \\
FRB230125D & 0.3265 & 640.08 & 88.0 & \cite{pastor2025localisation} \\
FRB230626A & 0.327  & 452.723 & 39.3 & \cite{sharma2024preferential} \\
FRB180301A & 0.330400 & 552.00 & 151.7 & \cite{price2019fast} \\
FRB231220A & 0.335500 & 491.20 & 49.9 & \cite{connor2024gas} \\
FRB211203C & 0.343900 & 636.20 & 63.7 & \cite{shannon2025commensal} \\
FRB230808F & 0.347200 & 653.20 & 27.0 & \cite{hanmer2025contemporaneous} \\
FRB220208A & 0.351000 & 437.00 & 101.6 & \cite{sharma2024preferential} \\
FRB230902A & 0.361900 & 440.10 & 34.1 & \cite{shannon2025commensal} \\
FRB220726A & 0.361900 & 686.23 & 89.5 & \cite{sharma2024preferential} \\
FRB220717A & 0.362950 & 637.34 & 118.3 & \cite{rajwade2024study} \\
FRB200906A & 0.3688 & 577.8 & 35.8 & \cite{bhandari2022characterizing} \\
FRB220330D & 0.3714 & 468.1 & 38.6 & \cite{sharma2024preferential} \\
FRB240119A & 0.376000 & 483.10 & 38.0 & \cite{connor2024gas} \\
FRB190611B & 0.377800 & 321.40 & 57.8 & \cite{heintz2020host} \\
FRB220501C & 0.381000 & 449.50 & 30.6 & \cite{shannon2025commensal} \\
FRB230613A & 0.392300 & 483.51 & 30.0 & \cite{pastor2025localisation} \\
FRB220204A & 0.401200 & 612.58 & 50.7 & \cite{connor2025gas} \\
FRB230712A & 0.452500 & 587.57 & 39.2 & \cite{sharma2024preferential} \\
FRB230907D & 0.463800 & 1030.79 & 29.0 & \cite{pastor2025localisation} \\
FRB181112A & 0.475500 & 589.27 & 102.0 & \cite{prochaska2019low} \\
FRB231020B & 0.477500 & 952.20 & 34.0 & \cite{pastor2025localisation} \\
FRB220307B & 0.248123 & 499.27 & 128.2 & \cite{law2024deep} \\
FRB220310F & 0.477958 & 462.24 & 46.3 & \cite{law2024deep} \\
FRB220918A & 0.491000 & 656.80 & 41.0 & \cite{shannon2025commensal} \\
FRB231210F & 0.500000 & 720.60 & 32.0 & \cite{pastor2025localisation} \\
FRB190711A & 0.522000 & 593.10 & 56.5 & \cite{heintz2020host} \\
FRB230216A & 0.531000 & 828.00 & 38.5 & \cite{sharma2024preferential} \\
FRB221219A & 0.553000 & 706.71 & 44.4 & \cite{sharma2024preferential} \\
FRB230814A & 0.553000 & 696.40 & 104.8 & \cite{connor2024gas} \\
FRB190614D & 0.600000 & 959.20 & 87.8 & \cite{law2024deep} \\
FRB210924D & 0.600000 & 737.00 & 45.0 & \cite{carli2024trapum} \\
FRB231010A & 0.610000 & 442.59 & 41.0 & \cite{pastor2025localisation} \\
FRB220418A & 0.622000 & 623.25 & 36.7 & \cite{law2024deep} \\
FRB220224C & 0.627100 & 1140.20 & 52.0 & \cite{sharma2024preferential} \\
FRB190523A & 0.660000 & 760.80 & 37.2 & \cite{ravi2019fast} \\
FRB190208A & 0.830000 & 580.01 & 71.5 & \cite{hewitt2024repeating} \\
FRB220222C & 0.853000 & 1071.20 & 56.0 & \cite{pastor2025localisation} \\
FRB240123A & 0.968000 & 1462.00 & 90.2 & \cite{connor2024gas} \\
FRB221029A & 0.975000 & 1391.75 & 43.8 & \cite{sharma2024preferential} \\
FRB220610A & 1.016000 & 1458.15 & 31.0 & \cite{gordon2024fast} \\
FRB230521B & 1.354000 & 1342.90 & 138.8 & \cite{connor2024gas} \\
FRB240304B & 2.148000 & 2330.00 & 28.1 & \cite{caleb2025fast} \\
\hline
\end{longtable}
\endgroup
\twocolumngrid

\bibliographystyle{apsrev4-1}
\bibliography{frb_bib.bib}

\end{document}